\documentstyle[twoside,fleqn,espcrc2,epsf]{article}

\newcommand{\AmS}{{\protect\the\textfont2
  A\kern-.1667em\lower.5ex\hbox{M}\kern-.125RMS}}

\newcommand{\rem}[1]{{}}
\newcommand{\vev}[1]{\langle#1\rangle}
\def\slash{\!\!\!/}
\def\rarr{\rightarrow}

\def\NN{{\cal N}}
\def\order{{\cal O}}

\def\none{$\NN=1$}
\def\ntwo{$\NN=2$}

\def\nfour{$\NN=4$}
\def\susy{supersymmetry}
\def\susic{supersymmetric}
\def\Lgr{{\cal L}}
\def\gt{gauge theory}
\def\nc{N_c}
\def\nf{N_f}
\def\lam{\lambda}
\def\Lam{\Lambda}
\def\al{\alpha}
\def\mn{{\mu\nu}}
\def\ias{School of Natural Sciences,
Institute for Advanced Study,
Olden Lane,
Princeton, NJ 08540, USA}
\def\ZZ{{\bf Z}}
\def\ZN{\ZZ_N}

\newcommand{\be}[1]{\begin{equation}\label{#1}}
\newcommand{\ee}{\end{equation}}
\newcommand{\FFig}[1]{Fig.~{\ref{fig:#1}}}
\newcommand{\eref}[1]{(\ref{#1})}

\title{
IASSNS--HEP--98/91\\
hep-ph/9810059 \\ 
~ \\
QCD, Supersymmetric QCD, Lattice QCD and String Theory:\\
Synthesis on the Horizon?}

\author{M.J. Strassler\address{\ias}\thanks{Research supported in
        part by National Science Foundation grant NSF PHY-9513835 and
        by the W.M.~Keck Foundation. } }
       
\begin{document}

\begin{abstract}
Supersymmetric gauge theories in four dimensions have taught us many
important physics lessons.  These can both inform and be informed by
future work on the lattice.  I focus on three issues: the properties
of supersymmetric Yang-Mills theory and its relation to the
non-supersymmetric case; the properties of gauge theories with matter
and their relation to real QCD; and, briefly, the recent discovery
that gauge theories and string theories are more deeply connected than
ever previously realized.  Specific questions for lattice gauge
theorists to consider are raised in the context of the first two
topics.
\end{abstract}

\maketitle

\section{Introduction}

The field of supersymmetric gauge theory is vast, fascinating, and
technical, and obviously cannot be reviewed in an hour. My limited
intention today is to give a qualitative and conceptual talk, with an
eye toward conveying what I see as the most important messages for
lattice gauge theory arising from the recent advances in this branch
of mathematical physics.  Consequently, this lecture will consist
merely of results, largely those of other authors; it will neither
contain descriptions of the evidence for these results nor any
technical details, and referencing will be limited. I learned much of
the physics in conversations with Seiberg and with my collaborators
Intriligator and Leigh.  However, the synthesis presented here has not
been emphasized elsewhere.

The purpose of my giving such a lecture to an audience of lattice
gauge theorists?  This is perhaps conveyed best by Tom DeGrand's
words: he requested that I encourage you ``to change the line in
[your] code that sets $N_c,N_f=3$.'' I hope to convince you that a
systematic exploration of theories {\it other} than real-world QCD is
an important and exciting direction for research, and that placing QCD
in the context of a wider variety of theories may become a powerful
tool for understanding it.

The topics I will cover today are the following. First, I will discuss
\none\ super-Yang-Mills (SYM) theory and compare it with Yang-Mills
(YM) theory and with QCD (by which I mean YM with matter --- unless
otherwise specified, fermions in the fundamental representation.)
Second, I will discuss SQCD (\none\ SYM with fermions and scalars in
the fundamental representation) and its relation to QCD.  And finally,
I will briefly discuss the recently discovered connection relating
gauge theory to gravity and string theory.  (I have also written
lectures for non-experts in \cite{ykis} and \cite{nonBPS}; parts of
the present talk have been abridged because of overlap with the
previous articles.)

\subsection{Beyond QCD}

There are several reasons why physicists should seek a detailed
understanding of gauge theories beyond real-world QCD.  First, we can
use them to gain insight into the properties of the real world.  It
would be helpful to know which aspects of the strong interactions are
special to $\nc=\nf=3$ and which ones are generic, or at least common
to many models. Second, a theory with behavior only vaguely similar to
QCD may be responsible for electroweak symmetry breaking, as in
technicolor and topcolor models.  Examples of non-QCD-like theories
are the fixed-point models discussed in \cite{pmtechni}.  Third, it is
important to test numerically some of the analytic predictions of
supersymmetric gauge theory, in part to close some remaining loopholes
in the analytic arguments.  And finally, there are possible
applications to condensed matter.

Four-dimensional supersymmetric gauge theories are good toy models for
non-supersymmetric QCD and its extensions.  Some of these theories
display confinement; of these, some break chiral symmetry but others
do not \cite{nsexact,powerholo}.  The mechanism of confinement
occasionally can be understood using a weakly-coupled ``dual
description'' \cite{nsewone,DS,NAD} (an alternate set of variables for
describing the same physics.)  However, what is more striking is that
most of these theories do not confine \cite{NAD}!  Instead, their
infrared physics is governed by other, unfamiliar phases, often
described most easily using dual variables.  (Here and throughout,
``phase'' refers to the properties of the far infrared physics at zero
temperature.)  The phase diagram, as a function of the gauge group,
matter content, and interactions, is complex and intricate \cite{NAD}.
Recently, it has been found that the large-$\nc$ physics of these
theories may be tractable --- and that it is superstring theory
\cite{gubkleb,maldacon,GKP,ewAdS}!  I will discuss all of these
issues below.

\section{Pure \none\ Super-Yang-Mills}

Consider $SU(N)$ gauge theory with a vector boson $A_\mu$ and a
Majorana spinor $\lam_\al$, with Lagrangian
\be{noneLgr}
\Lgr = {1\over 2g^2} \left[{\rm tr}\ F_{\mu\nu}F^{\mu\nu} 
+ i\bar\lambda D\slash \lambda \right]
\ee
This is \none\ SYM theory.  It exhibits chiral symmetry breaking: a
gluino condensate $\vev{\lam\lam}$ forms, leading to $N$ equivalent
vacua, differing only by the phase of the condensate, which rotate
into one another under $\theta\rarr\theta+2\pi$.  Domain walls can
separate regions of different vacua.  The theory confines and exhibits
electric flux tubes.

\subsection{Chiral Symmetry Breaking}

Let us examine the issue of chiral symmetry breaking.  The classical
$U(1)$ axial symmetry $\lam\rarr\lam e^{i\sigma}$ is anomalous; since
$\lam$ has $2N$ zero modes in the presence of an instanton, only a
$\ZZ_{2N}$, under which $\lambda \rarr \lambda e^{i\pi n/N}$, is
anomaly-free.  The dynamics leads to $N$ equivalent vacua with
$\vev{\lam\lam}=\Lam^3 e^{i(2\pi n/N)}$, which breaks the $\ZZ_{2N}$ to a
$\ZZ_2$ under which $\lam\rightarrow -\lam$.  Note that an instanton
in this theory comes with a factor $\Lam^{3N}$; the form of the gluino
condensate suggests it is induced by $1/N$
fractional instantons.

\begin{figure}
\centering
\epsfxsize=1.0in
\hspace*{0in}\vspace*{0in}
\epsffile{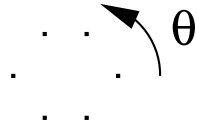}
\caption{The vacua of SYM theory are rotated
by the theta angle.}
\label{fig:vacua}
\end{figure}
\begin{figure}
\centering
\epsfxsize=2.3in
\hspace*{0in}\vspace*{0in}
\epsffile{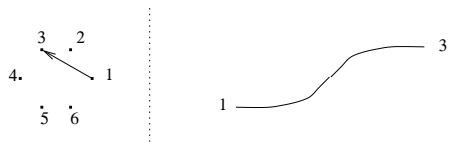}
\caption{Domain walls can interpolate between
vacua of SYM theory.}
\label{fig:walls}
\end{figure}

If we shift the theta angle by $\theta\rarr\theta+\alpha$, then
$\lambda\rarr\lambda e^{i\alpha/2N}$, and the $N$ equivalent vacua
(\FFig{vacua}) are rotated by an angle $\alpha/N$.  Any given vacuum
only comes back to itself under $\theta\rarr \theta + 2\pi N $, but
the physics is invariant under $\theta\rarr \theta + 2\pi $.  Domain
walls can exist between regions in different vacua; as in
\FFig{walls}, the condensate $\vev{\lambda\lambda}$ can change
continuously from $\Lambda^3e^{2\pi i n/N}$ to $\Lambda^3e^{2\pi i
n'/N}$.

It is important to note that similar properties --- equivalent
discrete vacua, domain walls, and possible fractional instantons ---
might occur (and should be searched for) in certain non-supersymmetric
gauge theories with fermions.

\subsection{Confinement and Flux Tubes}

[I have abbreviated some of the content of this section; the
interested reader can find more detail in \cite{ykis}.]

\none\ SYM is confining, with electric flux tubes which as always
carry charge in the center of the gauge group; for $SU(N)$ they carry
quantum numbers $k$ in $\ZZ_N$.  The tension $T_k$ of these confining
strings is a function of $k,N,\Lam$; note $T_k=T_{N-k},T_N=0$ by
$\ZZ_N$ symmetry.  An electric source of charge $k$ ({\it e.g.} one
in a $k$-index antisymmetric tensor representation) will be confined
by a $k$-string (a string carrying $k$ units of flux.)  The ratio
$T_k/T_{k'}$ is a basic property of YM and SYM, as fundamental as
the hadron spectrum and easier for theorists to estimate.  Two
interesting predictions for this ratio are obtained through the
Hamiltonian strong coupling expansion, which gives the second Casimir
of the representation
\be{strongTk}
T_k \propto k(N-k)/N   
\ee 
and through weakly broken \ntwo\ \susic\ \gt\ \cite{DS}
\be{DSTk}
T_k \propto \sin {\pi k \over N}
\ee 
A question of considerable interest is whether either formula well describes
SYM and/or YM theory.  

A more qualitative question is whether YM/SYM is a Type I or Type II
dual superconductor --- that is, whether $T_2<2T_1$ or $T_2>2T_1$.  In
the former case flux tubes attract one another, in the other they
repel.  On general grounds I personally expect that $2T_1>T_2>T_1$ for
$SU(N)$, $N>4$.  The reasoning for this is the following.  For $N=2$,
we have $T_2=0$, while for $N=3$ we have $T_2=T_1,T_3=0$ on general
grounds.  For large $N$, we would expect $T_k= kT_1 \pm \order(1/N)$
since a $k$-string is a bound (or unbound) state of $k$ $1$-strings,
but string-string interactions are of order $1/N$.  Any reasonable
interpolating formula will satisfy $kT_1>T_k>T_1$ for $N>4$,
$1<k<N-1$. Ohta and Wingate \cite{OW} have compared $T_1$ and $T_2$ in
$SU(4)$ by computing the potential energy between sources in the ${\bf
4}$ and ${\bf\bar4}$ representation ($V_{4\bar4}(r) \sim T_1r$) and the
energy between sources in the ${\bf 6}$ representation ($V_{66}(r)
\sim T_2 r$.)  Their preliminary results indicate indeed that
$2T_1>T_2>T_1$.

\subsection{Mechanism of Confinement}

[The material of this section was carefully covered in \cite{ykis},
and has been heavily abridged.  I have attempted here only to
outline the results, and focus attention on the physics message.]

What drives confinement?  This question cannot be addressed directly
in SYM, because this theory does not have a weakly-coupled dual
description. However, there is a trick for studying this question ---
within limits, as discussed further below.  The trick is the
following.  One can add additional massive matter to SYM without
leaving its universality class (note supersymmetry and holomorphy
ensure this; see \cite{powerholo}) In particular I will study broken
\ntwo\ SYM \cite{nsewone,DS} and broken \nfour\ SYM
\cite{cvew,rdew,ykis} gauge theories which have the same massless
fields as \none\ SYM.  These theories have a duality transformation
which allows their dynamics to be studied using a weakly coupled
magnetic description.  In this description it will be possible to show
that there are monopoles in the theory, which condense, thereby
causing confinement \cite{nsewone}.  The required $\ZZ_N$ flux
tubes will appear as string solitons.  The picture which emerges will
strongly support the old lore on the Dual Meissner effect.

However, a strong word of caution is in order here.  In particular,
although these theories are in the same universality class as \none\
SYM, they are not equivalent to it.  While confinement and an energy
gap are universal properties of all of these theories, the monopoles
which lead to confinement are {\it not} universal.  The properties of
the monopoles will depend on the matter that is added to SYM.  We will
shortly see that this poses problems for abelian projection approaches
to confinement.

I will now illustrate these points by studying the broken \ntwo\ and
\nfour\ SYM theories.  The \ntwo\ $SU(N)$ SYM theory, with strong
coupling scale $\Lam$, has a dual description as an abelian
$U(1)^{N-1}$ \gt \cite{nsewone,sun,suntwo,sunthree}; the monopoles of
the $SU(N)$ description are electrically charged under the dual
description.  When masses $\mu\ll\Lam$ are added so that the only
massless fields are those of \none\ SYM, the monopoles develop
expectation values.  The dual description of this condensation
involves nothing more than $N-1$ copies of the Abelian Higgs model,
and so gives $N-1$ solitonic Nielsen-Olesen strings \cite{nsewone,DS},
each carrying an integer charge.  Electrically charged sources are
therefore confined, and the flux tubes which confine them are the
solitonic strings of the dual description.  However, the confining
strings are problematic \cite{DS,hsz}.  Although they carry an exact $\ZZ_N$
symmetry, they also each carry an (approximate) integer charge,
violated only at the scale $\Lam$ which is large compared to the
string tension.  This extra symmetry leads the theory to exhibit not
one but $N/2$ Regge trajectories --- thus the theory does not have the
same properties as YM or \none\ SYM.  As $\mu\rarr\Lam$, the extra
symmetry begins to disappear, but at the same time the magnetic
description becomes strongly coupled, so no reliable dual description
can be given in the regime where the theory behaves as \none\ SYM is
expected to do.

\begin{figure}
\centering
\epsfxsize=2.5in
\hspace*{0in}\vspace*{0in}
\epsffile{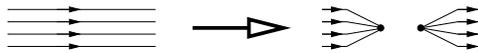}
\caption{The annihilation of $N$ flux tubes using baryon vertices;
this transition would be inhibited if a good dual description of
YM theory involved $U(1)^{N-1}$.}
\label{fig:noannih}
\end{figure}

Note these properties are characteristic of abelian projection
approaches to confinement.  If we project $SU(N)\rarr U(1)^{N-1}$,
dynamically or otherwise, then abelian monopole condensation leads to
Nielsen-Olesen strings, each with its own approximately conserved
integer charge.  This unavoidable charge inhibits annihilation of $N$
identical strings (as in \FFig{noannih}) which both leads too an
overabundance of metastable hadrons and to difficulty in forming
baryons \cite{hsz}.  This is a serious problem for abelian projection
approaches to QCD.

The situation in broken \nfour\ SYM \cite{ykis} is much more
satisfactory.  \nfour\ $SU(N)$ SYM is a conformal field theory (CFT), with
gauge coupling $g$.  Its magnetic description, also an \nfour\
conformal field theory, has gauge group $SU(N)/\ZN$ and coupling
$1/g$; thus, if $g\gg 1$, the magnetic description is weakly coupled.
Adding masses $\mu$ to all but the fields of \none\ SYM causes the
scalars of the magnetic description to condense, breaking the dual
gauge group completely \cite{cvew,rdew}.  This {\it non}-Abelian Higgs
model has string solitons, but because the fundamental group of
$SU(N)/\ZN$ is $Z_N$, these strings carry a $\ZN$ charge \cite{ykis},
in contrast to the integer charges found in the case of broken \ntwo\
SYM. In short, the electric description of the theory involves
confinement by electric flux tubes carrying $\ZN$ electric flux,
leading to a single Regge trajectory, in agreement with expectations
for \none\ SYM.  Notice that the associated dual description does not
resemble abelian projection: it is essential for the $\ZN$ charges of
the strings that the dual theory is {\it non-abelian}.  Since one
cannot obtain the dual $SU(N)/\ZN$ theory by a projection on the
$SU(N)$ theory --- the relation between the two is fundamentally
quantum mechanical --- it seems to me that abelian projection is
disfavored.

However, one cannot carry this logic all the way to the \none\ SYM
theory itself. To do so requires taking $\mu\rarr\infty,g\rarr 0$, but
in this limit the magnetic theory becomes strongly coupled and the
semiclassical discussion of the previous paragraph becomes unreliable.

What has been obtained here?  Two qualitatively different descriptions
of confinement have been found, and neither can be continued directly
to the theory of interest.  One looks similar to abelian projection,
while the other absolutely does not.  What are we to make of this
situation, and how are we to reconcile apparent contradictions?

\begin{figure}
\centering
\epsfxsize=2.5in
\hspace*{0in}\vspace*{0in}
\epsffile{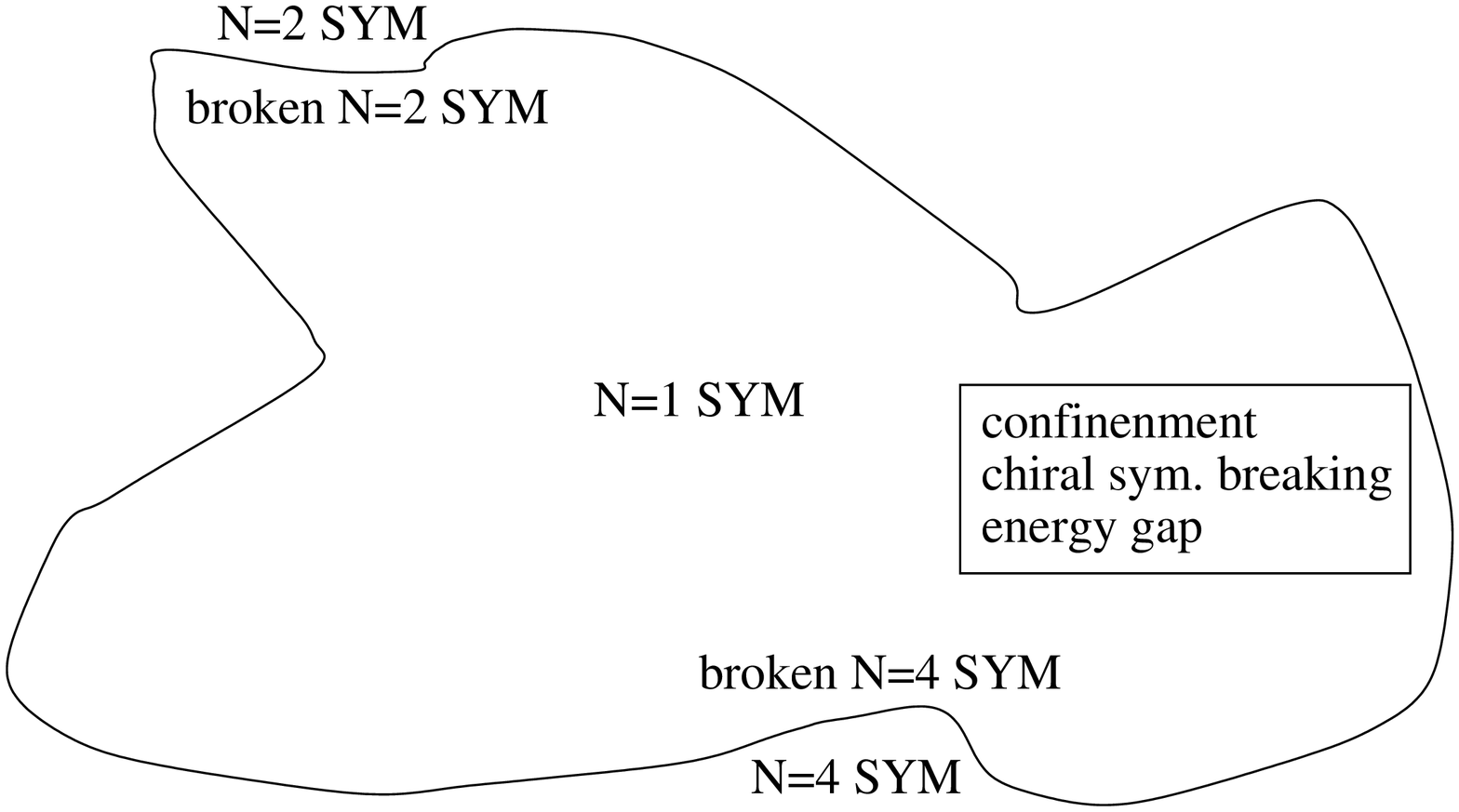}
\caption{The universality class containing \none\ SYM,
with \ntwo,\nfour\ SYM lying just outside.}
\label{fig:thyspace}
\end{figure}

I believe the correct way to view this situation\footnote{I thank
N. Seiberg for discussions on this point.} is the following.  Consider
the space of theories in the same universality class as SYM,
\FFig{thyspace}.  Although all of these have a gap, confinement, and
chiral symmetry breaking, only theories near a phase boundary, at the
edge of the space, may be expected to have a weakly-coupled
Landau-Ginsburg-type description.  These dual descriptions may be used
to establish the universal properties of \none\ SYM.  Theories far
from the boundary, such as \none\ SYM itself, may simply not have any
such description, and so there may not be any weakly-coupled effective
theory for describing the string charges, hadron spectrum, and
confinement mechanism of SYM.  The same logic may apply to non-\susic\
YM.

In the end, then, our conclusions are very weak --- we can show
SYM is confining, but we cannot really say much about the mechanism
which confines it.  Any monopole description of
confinement in SYM or YM is likely to be strongly coupled.  Is such a
description useful? or unambiguous?  It seems unlikely to be
predictive, in any case. It may be disappointing, but it appears
likely there is no simple magnetic description of confinement in
nature.

\subsection{Breaking Supersymmetry}

What happens if we break \susy\ by adding a mass $m$ to the gaugino?

\be{gauginomassLgr}
\Lgr \rarr \Lgr + m \lambda\lambda
\ee
The degeneracy of the $N$ vacua will be broken, and the one with
lowest energy will depend on the phase of $m$.  As
$\theta\rarr\theta+2\pi$, the preferred vacuum will shift from one to
the next. All domain walls become unstable (except for
$\theta=(2n+1)\pi$, where the two lowest vacua become degenerate.)
However, the energy gap, the spectrum, and the general features of
confinement will not be altered if $|m|\ll|\Lambda|$.

A question for lattice study is whether there are any qualitative
transitions in the physics of the theory as $m$ increases and pure YM
is recovered.  For example, consider $T_k(m)$; does it change
continuously as a function of $m$? Are the confining flux tubes
essentially similar in YM and SYM?  It would also be interesting to
understand the behavior of $\vev{\lam\lam}$ as a function of $m$.  If
no dramatic transitions are seen, then the issues discussed earlier in
regard to SYM apply also to YM, in fact to broken \none\ SYM for all
values of $m$. 

\subsection{Linkage of YM to \nfour\ SYM}

Most physicists outside of the field of \susy\ are inclined to think
of supersymmetric theories, especially those with extended
supersymmetry, as esoteric curiosities with no real importance for
physics.  I now intend to convince you that this is far from the
truth, and that, in fact, there is a direct link between the
spectacular properties of \nfour\ SYM and the properties of ordinary
non-\susic\ YM.

\begin{figure}
\centering
\epsfxsize=2.7in
\hspace*{0in}\vspace*{0in}
\epsffile{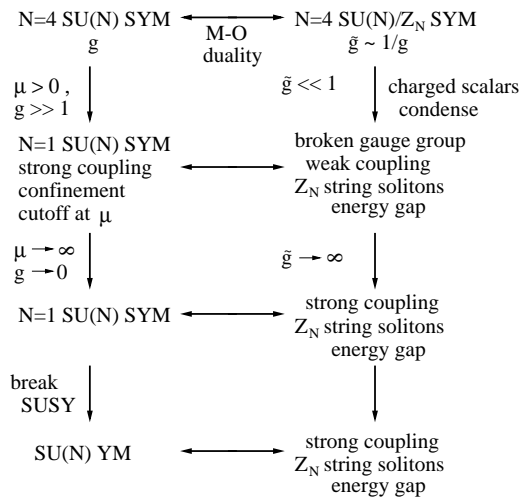}
\caption{The connection of Montonen-Olive duality
to confinement in Yang-Mills theory.}
\label{fig:linkageA}
\end{figure}

Consider the linkage diagram in \FFig{linkageA}.  We begin at the top
with \nfour\ $SU(N)$ SYM theory, a conformal field theory with a gauge
coupling $g$. Under Montonen-Olive duality, this theory is rewritten,
using magnetic variables, as \nfour\ $SU(N)/\ZN$ SYM, with gauge
coupling $\tilde g\propto 1/g$.  Next, as discussed earlier, we take
$g\gg 1$ and break \nfour\ \susy\ to \none\ by adding finite masses
$\mu$ for some of the fields.  The resulting theory is strongly
coupled, confines, and has chiral symmetry breaking, as is easily seen
using the weakly-coupled magnetic variables, where the effect of the
\susy\ breaking is to cause the light monopoles to condense, breaking
the gauge group completely, and leading to $\ZN$-carrying solitonic
flux tubes.

The following step is to take $g$ small and $\mu$ to infinity, holding
the strong coupling scale fixed.  The theory becomes pure \none\ SYM
in this limit, and maintains confinement and chiral symmetry breaking
since it remains in the same universality class.\footnote{We know this
because the theory is holomorphic in the parameter $\mu$; see
\cite{powerholo}.}  The magnetic variables become strongly coupled as
$g\rarr 0$, but we expect the $\ZN$ solitonic strings will survive,
since their existence is a consequence of the topology of the
$SU(N)/\ZN$ gauge group.

The last step is to break \none\ SYM to YM.  Here I must assume that
there are no sharp transitions between these two theories --- an issue
which can be addressed on the lattice, as I discussed above --- in
order to make the linkage complete.  However, because SYM and YM share
many properties, such a conjecture is quite plausible.  If the
transition between these two theories is smooth, then confinement and
the $\ZN$ flux tubes will survive from one theory to the other.

In summary, modulo the conjecture that \none\ SYM and YM are
continuously connected, the specific structure of duality in \nfour\
SYM theories is directly related to --- perhaps even implies --- the
fact that YM is a confining theory with $\ZN$ flux tubes.

\section{Gauge Theories with Matter}

Supersymmetric gauge theories teach us that the matter content of a theory
plays an essential role in determining its basic physics.  In
particular, the phase at zero temperature of a \gt\ (that is, its
properties in the far infared) depends in a complicated way on (1) its
gauge group $G$, (2) its matter representations $R$, and (3) the
self-interactions $\Lgr_m$ of the matter, including non-renormalizable
interactions. Recent work has shown that the phase structure of \none\
\susic\ theories is complex and intricate, and it is reasonable to
expect that the same will be true of non-\susic\ theories.

Among the surprises discovered in the \susic\ context are that there
are new and unexpected phases unknown in nature or previously thought to be quite
rare.  It also appears that duality is a fundamental property of field
theory (and also of gravity, and even between gravity and gauge
theory!)  Certain accepted or at least popular lore has been refuted
as well: the beta function does {\it not}, by itself anyway, determine
the phase of a gauge theory; confinement does not always cause chiral
symmetry breaking; the 't Hooft anomaly matching argument in favor
of such chiral symmetry breaking in strongly coupled theories can be
evaded; and fixed points in four dimensions are not at all rare --- in
fact, they are commonplace!

\subsection{SQCD}

\begin{figure}
\centering
\epsfxsize=2.7in
\hspace*{0in}\vspace*{0in}
\epsffile{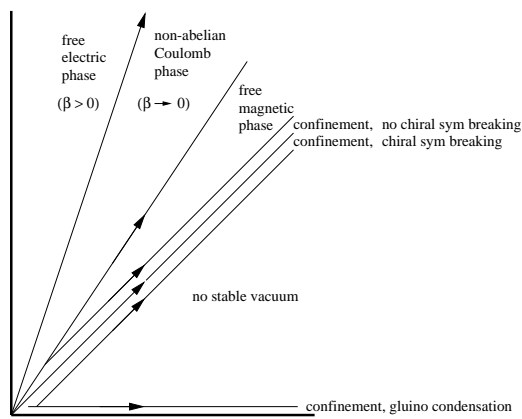}
\caption{The phases of SQCD.}
\label{fig:phases}
\end{figure}

Supersymmetric QCD consists of \none\ $SU(N)$ SYM along with $\nf$
flavors of massless scalars (squarks) and fermions (quarks) in the
fundamental representation.  \FFig{phases} shows the phase diagram of the
infrared behavior of the theory as a function of $N$ and $\nf$,
as determined by Seiberg in \cite{NAD}. 

When $\nf>3N$, the theory has a positive beta function and flows to
weak coupling.  Since the theory is free in the infrared in terms of
the original variables, this is called the free electric phase.

When $3N>N_f>{3\over2}N$, the theory flows toward strong coupling, but
the coupling eventually stops running.  The low-energy theory is an
interacting conformal fixed point.  This is called the ``non-abelian
Coulomb phase''.  The theory has a ``dual description'' using
``magnetic'' variables.

When ${3\over2}N\geq \nf\geq N+2$, the theory is in the ``free
magnetic phase''.  The theory flows to strong coupling in the
infrared, but the dual variables become weakly coupled.  The theory of
the dual variables is an infrared-free $SU(\nf-N)$ gauge theory.

For $N_f=N+1$, the theory undergoes ``confinement without chiral
symmetry breaking''.  The light particles are massless mesons and
(scalar) baryons, and the theory describing them is a linear sigma
model.  For $N_f=N$, the theory displays ``confinement with chiral
symmetry breaking''.  Again the light particles are massless mesons
and (scalar) baryons, but now the theory describing them breaks chiral
symmetry and becomes a nonlinear sigma model.\footnote{In both these
cases the term ``confinement'' should be taken with a grain of salt,
as $SU(N)$ with fundamentals has no phase boundary between confining
and Higgs phases, and all flux tubes break due to pair production;
however in certain other models confinement is unambiguous, and even
there chiral symmetry is not always broken.}

When $\nf=N-1$, instantons generate an unstable potential for the
squarks; a similar effect, due to gaugino condensation (or perhaps
fractional instantons?) occurs for $N-2\geq \nf\geq 1$.  For $\nf=0$,
as discussed above, we have confinement and gaugino condensation.

\subsection{Non-Abelian Coulomb Phase}

\begin{figure}
\centering
\epsfxsize=2.9in
\hspace*{0in}\vspace*{0in}
\epsffile{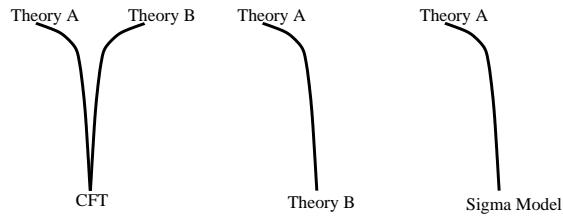}
\caption{(a) Two gauge theories flow to the same conformal fixed
point. (b) Gauge theory A becomes strongly coupled, and at low-energy
mysteriously turns into gauge theory B.  (c) Gauge theory A becomes
strongly coupled, and mysteriously turns into a theory of mesons and
baryons.}
\label{fig:flow}
\end{figure}

The conformal field theories which are found in this phase are known
to exist in perturbation theory at large $N$ and $\nf$.  These
perturbative fixed points, often called Banks-Zaks fixed points
although they predate those authors, are found both in SQCD and in
QCD.  We now know that in the case of SQCD the fixed points are found
far from large $N$ and $\nf$.  These fixed points exhibit duality:
there exist multiple gauge theories which flow to the same conformal
field theory, as illustrated schematically in \FFig{flow}a, and
thus multiple sets of variables by which the conformal field theory
may be described.  This is analogous to Montonen-Olive duality in
\nfour\ SYM, which is a conformal field theory.

\subsection{Free Magnetic Phase}

This phase was entirely unexpected.  Here, the theory flows to strong
coupling, and its infrared physics is described by weakly coupled
composite matter and gauge fields.  These composites are non-local
with respect to the original degrees of freedom, and have an unrelated
gauge symmetry.  The duality transformation which acts on the fixed
points of the non-abelian Coulomb phase operates in the free magnetic
phase by exchanging the fundamental theory of the ultraviolet with the
infrared-free composite theory, as illustrated in \FFig{flow}b.  Note
that the free magnetic phase is {\it not} a confining phase, as proven
in \cite{spinmono,nonBPS}.

\subsection{Confinement with and without chiral symmetry breaking}

In the confining theories, the low-energy description is a theory of gauge
singlets built from polynomials in the original degrees of freedom.
The main difference between chiral-symmetry-preserving and -breaking
theories is in the interaction Lagrangian, which determines the
symmetries preserved by the vacuum.

As in the free magnetic phase, the duality transformation exchanges
the ultraviolet theory with the infrared one --- the quarks and gluons
of the gauge theory are exchanged with the mesons and baryons of the
linear or non-linear sigma model (\FFig{flow}c).  At this point duality
begins to resemble what is done in real-world QCD when we rewrite the
theory in term of hadrons and the chiral Lagrangian.  This strongly
suggests that the QCD/chiral-Lagrangian ``duality'' transformation is
conceptually related to electromagnetic duality, its generalization to
Montonen-Olive duality, and its lower-dimensional cousins.  I will
make this more precise below, using a second linkage diagram; see also
\cite{ykis} for more details.

\rem{
\subsection{Dependence of Phase on Gauge Group and Matter Content}
While the phase structure of different gauge theories roughly resemble
one another, they are not by any means identical.  Consider, for
example, $SO(8)$ gauge theory with six fields in the vector
representation.  This theory is in the free magnetic phase, with an
abelian dual gauge group.  If we instead take $SO(8)$ with five
vectors and one spinor --- this theory has the same one-loop beta
function as the previous one --- then the theory is confining, and
does not break its chiral symmetries (CHECK!)  As another example,
compare $SU(N)$ SQCD with $\nf=N$ to \ntwo\ $SU(N)$ SYM.  The two
theories have the same beta function, but the first is in the
confining phase with chiral symmetry breaking, while the second is in
the free magnetic phase.
\subsection{Dependence of Phase on Interactions}
The phase of a theory also depends strongly on the interactions
between the matter fields.  For example, consider $SU(4)$ SQCD with
$\nf$ flavors of squarks $Q_i$, antisquarks $\tilde Q^i$, quarks and
antiquarks $\psi_i,\tilde \psi^i$.  Consider taking a
``superpotential'' $W=Q_1Q_2Q_3Q_4$ and adding the following
dimension-six perturbation to the Lagrangian:
\be{dimsix}
{1\over M^2}\left[
\sum_{i,j} {\partial^2 W\over \partial Q_i\partial Q_j}\psi_i\psi_j
+ h.c.
+\sum_i \left|{\partial W\over \partial Q_i}\right|^2
\right]
\ee
Although this perturbation is irrelevant at weak coupling, it can be
important in the infrared, depending on $\nf$.  In particular,
\begin{itemize}
\item
For $\nf=8$, the theory is in the non-abelian coulomb phasethe
perturbation \eref{dimsix} is irrelevant, as in the classical limit.
\item
For $\nf=7$, the theory without the perturbation is in the non-abelian
coulomb phase, but the perturbation is relevant and drives the theory
to a different conformal field theory; thus the phase is unchanged but
the particular fixed point is different.  
\item
For $\nf=6$, the theory
without the perturbation is in the free magnetic phase, but the
perturbation drives the theory to an interacting fixed point and thus
into the non-abelian coulomb phase.  
\item
For $\nf=5$, the theory is
confining; the perturbation obviously breaks some chiral symmetries,
but even more are broken dynamically as a result of the perturbation.
\end{itemize}
}

\subsection{Linkage of QCD to Duality in \ntwo, \none\ SQCD}

\begin{figure}
\centering
\epsfxsize=2.8in
\hspace*{0in}\vspace*{0in}
\epsffile{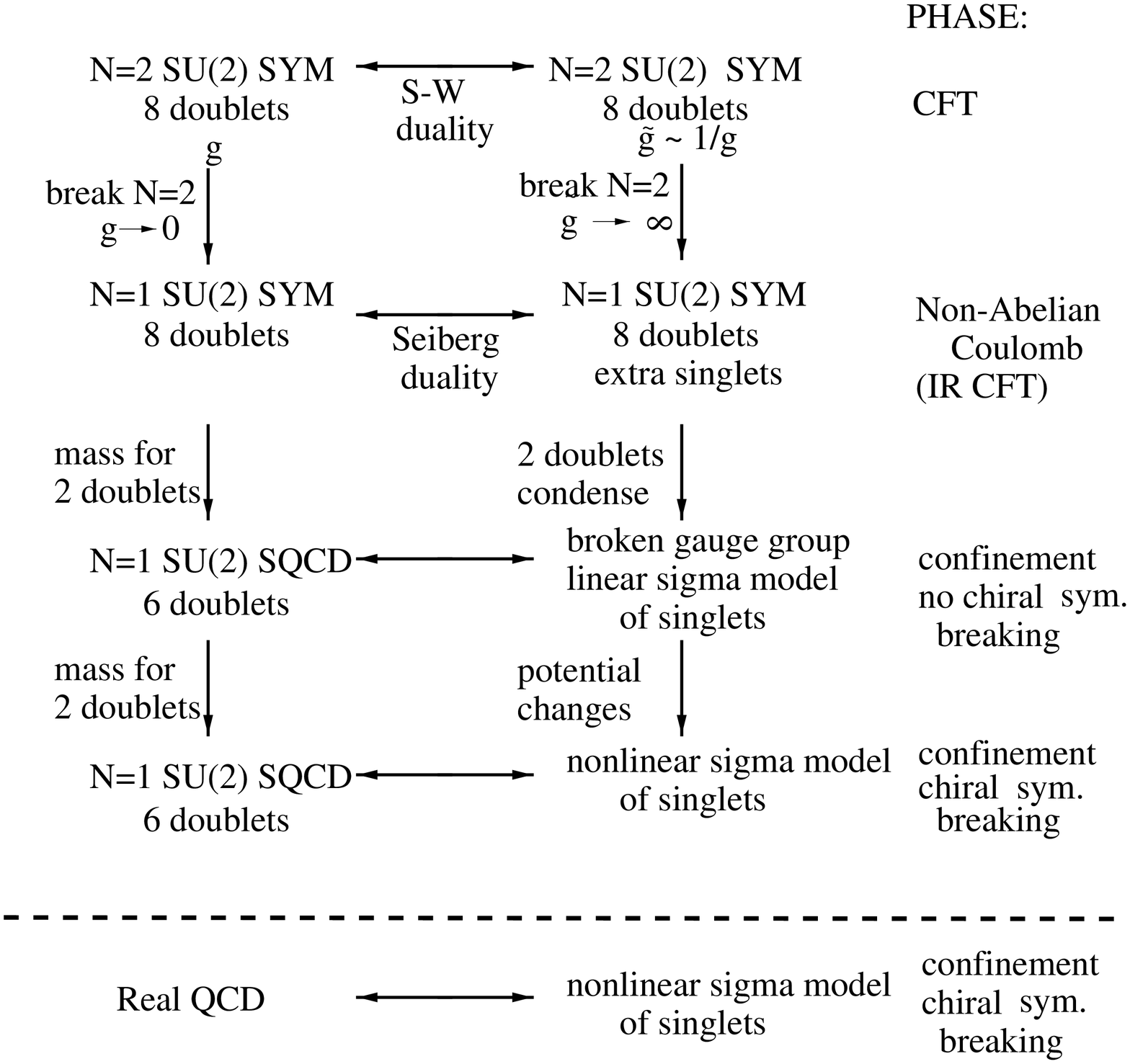}
\caption{The connection of Seiberg-Witten and Seiberg duality to
confinement, chiral symmetry breaking, and non-linear sigma models in
QCD.}
\label{fig:linkageB}
\end{figure}

Now I turn to my second linkage diagram, \FFig{linkageB}, which
relates the duality of finite \ntwo\ theories to that of \none\
theories and then to the properties of real QCD.

At the top of the diagram, we have a finite --- and therefore
conformal --- \ntwo\ SQCD theory, with gauge group $SU(2)$, gauge
coupling $g$, and eight quarks and squarks in the doublet
representation.\footnote{The choice of $SU(2)$ is for simplicity only;
the same physics applies with slight modification for any $SU(N)$.}
As in the case of \nfour\ SYM discussed earlier, this theory has a
representation in terms of magnetic variables as another \ntwo\ SQCD
theory, which in this case has the same gauge and matter content as
the electric variables but has coupling constant $1/g$.  This duality
transformation was discovered by Seiberg and Witten in their famous
paper of 1994 \cite{nsewtwo}.

Now let us break \ntwo\ \susy\ to \none\ by giving mass to the extra
fields in the \ntwo\ gauge multiplet.  The theory becomes \none\ SQCD
with gauge group $SU(2)$ and eight quarks and squarks in the doublet
representation.  This theory has a running coupling, but flows to a
conformal fixed point in the infrared --- it is in the non-abelian coulomb
phase.  As shown in \cite{emop}, the breaking of
\ntwo\ \susy\ causes the magnetic theory to flow to an \none\ SQCD
theory with the same charged matter content but with extra gauge
singlets and interactions, precisely those required by Seiberg's
\none\ duality transformation \cite{NAD}.  In other words, the
Seiberg-Witten duality transformation of the \ntwo\ theory flows to
the Seiberg duality transformation of \none\ SQCD.

Now add masses for two of the electric doublets, leaving a theory of
$SU(2)$ with six doublets.  This causes some magnetic squarks to
condense, breaking the $SU(2)$ magnetic gauge symmetry and leaving a
theory of massless gauge singlet fields $M$. These singlets are
precisely the mesons of the electric variables.  Since the magnetic
gauge symmetry is broken, electric charge is confined.  Thus,
confinement proceeds through a non-abelian generalization of the dual
Meissner effect, and the low-energy magnetic theory --- an
infrared-free non-renormalizable theory with a cutoff --- is the sigma
model describing the confined hadrons.  Examination of the sigma
model, in particular the potential energy $V(M)$, shows that the the
theory has a vacuum in which chiral symmetry is unbroken.

Adding masses for two more doublets merely causes the potential $V(M)$
to change in such a way that there is no longer a
chiral-symmetry-preserving vacuum.  Thus, $SU(2)$ SQCD with four
doublets confines and breaks chiral symmetry.  Shifting to the true
vacuum and renaming the fields as representing fluctuations around
that vacuum, we may rewrite the theory as a non-linear sigma model of
pions.

To go from here to real non-\susic\ QCD is a bit more of a stretch
than in the previous linkage diagram, because the removal of the
scalar squarks from the theory is rather more delicate and much less
well understood.  Rather than raise those questions, I leave the last
step in the diagram as more of a heuristic one.  It is evident that
the duality in \none\ $SU(2)$ SQCD with four doublets, relating the
gauge theory of gluons, quarks and their superpartners to a non-linear
sigma model of pions, is remarkably similar to the transformation
between real-world QCD and the chiral Lagrangian that we use to
describe its infrared physics.  Indeed it is completely justified to
call this QCD-to-hadron transformation ``duality''.  As we have seen,
the duality in confining SQCD can be derived from the Seiberg-Witten
duality of a conformal \ntwo\ SQCD theory. Is QCD-pion duality
likewise embedded in a chain of non-\susic\ duality transformations
similar to those in the diagram?

\subsection{Non-Supersymmetric Vectorlike Theories and the Lattice}

We are unable at this time to answer this question, even for
non-chiral theories like QCD, because almost nothing is known about
non-supersymmetric gauge theories other than $SU(2)$ and $SU(3)$ YM
and QCD with a small number of flavors.  This is where lattice
QCD comes in, as it is at the present time almost the only tool
available for studying this issue.

In \susic\ theories, we have begun to understand the complicated
subject of the long distance physics of gauge theories as a function
of their gauge group $G$, their matter representations $R_i$, and
their interaction Lagrangian (including non-renormalizable terms.)
Similar information would be welcome in the non-\susic\ case.  We know
that the non-abelian coulomb phase exists at large $\nc,\nf$ when the
one-loop beta function is very small, but how far does it extend away
from this regime?  What properties does the theory exhibit as it makes
the transition from the perturbative regime to the conformal regime?
The confining phase in supersymmetric theories is the exception, not
the rule; which non-supersymmetric theories actually confine?  Which
ones break chiral symmetry, and in what patterns?  What are their
confining string tensions $T_k$ and their low-lying hadron spectra?
Effects involving instantons, fractional instantons, monopoles, {\it
etc.} may play an important role in some theories --- but which ones,
and what effects are they responsible for?  The free magnetic phase
may not exist in non-supersymmetric theories --- perhaps it requires
the massless scalars of SQCD --- but it would be very exciting if it
were found (and a tremendous coup for the group which demonstrates its
existence!)  The existence of this phase would be a sufficient but not
necessary condition for gauge theory--gauge theory duality in the
non-\susic\ context, which could also perhaps be shown in the context
of the non-abelian coulomb phase.  And of course, we must not leave
out the possibility of new exotic effects which do not occur in the
\susic\ case.

It is important to emphasize that these questions are by no means of
purely academic interest.  The problem of electroweak symmetry
breaking has not been solved, and there remains the possibility that
the symmetry breaking occurs through a technicolor-like scenario, in
which it is driven by chiral symmetry breaking in a strongly-coupled
gauge theory.  Technicolor models with physics similar to QCD have
been ruled out by precision measurements at LEP.  However, if their
physics is considerably different from QCD, then we have no predictive
tools, and therefore no experimental constraints.  It will be an
embarassment to theorists if the LHC discovers evidence of strong
dynamics of a type that we simply do not recognize.  It is therefore
important for model building and for comparison with experiment that
we improve our understanding of strongly-coupled non-\susic\ theories.

I should mention also that there are potential spinoffs from such a
program in the areas of supersymmetry breaking (which can be detected
but not quantitatively studied using presently available analytic
techniques) and in condensed matter systems where many similar physics
issues arise.

To answer these questions, which require studying theories with very
light fermions and often a hiearachy of physical scales, will require
powerful lattice techniques not yet fully developed. I encourage you
all to think about how best to pursue this program of study.\footnote{A 
useful testing ground is to be found in three-dimensional abelian
gauge theories, both with and without supersymmetry.  The \susic\
theories are known to have an intricate phase structure, with a
duality transformation (called ``mirror symmetry'') and
interesting large $\nf$ behavior.  There are examples of non-trivial
fixed points, infrared-free mirror gauge theories, chiral symmetry
breaking and confinement.  Many of these phases are likely to show up
in the non-\susic\ case, and lattice approaches to studying non-\susic\
QCD could be tested in these theories.}

\section{Large $\nc$ Gauge Theory and String Theory}

The fact that gauge theory, in the limit of a large number of colors,
is in some way connected with string theory, with $1/N$ playing the
role of the string coupling, was first noted twenty-five years ago by
't Hooft.  While many have attempted to make progress in
understanding gauge theories by studying this limit further, a
quantitative approach to $SU(N)$ YM or QCD using $1/N$ as an expansion
parameter has been stymied by the difficulty of determining the
classical string theory which should appear in the $N\rarr\infty$
limit.

Let us consider some obvious facts about this string theory, and the
gauge theories which might be studied using this approach.  First,
critical (that is, $d=10$) superstrings have gravity and
supersymmetry.  Since QCD has neither light spin-two hadrons nor
supersymmetry, clearly its string theory is unrelated to critical
superstrings.  A second obvious fact is that only confining theories
have physical string-like flux tubes and associated area laws for
Wilson loops.  Other theories, including conformal field theories such
as \nfour\ SYM, have no physical string-like behavior and should not
be associated with string theories.

We will not get anywhere using these obvious facts, however, because
they are wrong.  Following on the work of Gubser and Klebanov
\cite{gubkleb}, and followed in turn by work of those authors with
Polyakov \cite{GKP}, of Witten \cite{ewAdS}, and then of a flood of
others, Maldacena proposed a bold conjecture \cite{maldacon} relating
supersymmetric conformal field theories in four dimensions to
superstring theory.  I will now state this conjecture, defining my
terms as I go.

\subsection{Maldacena's Conjecture}

According to Maldacena's idea, \nfour\ SYM theory with $N$ colors and
coupling $g$ is related to Type IIB superstring theory (a
ten-dimensional theory of closed strings with IIB supergravity as its
low-energy limits --- its massless fields are a graviton $G_\mn$,
antisymmetric tensor $B_\mn$ and dilaton $\phi$ along with
``Ramond-Ramond'' 0-index, 2-index and 4-index antisymmetric tensor
fields $\chi, A_\mn, C_{\mu\nu\rho\sigma}$.)  The string theory exists
on a ten-dimensional space consisting of a five-sphere
$(x^2_1+x^2_2+x^2_3+x^2_4+x^2_5+x^2_6=R^2$, a space of constant
positive curvature) times a five-dimensional Anti de Sitter space
$(-x^2_1-x^2_2+x^2_3+x^2_4+x^2_5+x^2_6=R^2$, a space of constant
negative curvature) with non-zero flux of $\partial_\kappa
C_{\mu\nu\rho\sigma}$.  The radius of the sphere and of the AdS space
are both $R=g^2N$, so the curvature of the space is small at large
$g^2N$, while the string coupling $g_s$ is the square of the gauge
coupling $g^2=R/N$.  (Notice the $N$ dependence accords with 't
Hooft's original observation.)

 Where is the four-dimensional gauge theory in this ten-dimensional
string?  The answer is fascinating.  The five-dimensional AdS space
has a boundary, an infinite spacelike distance away but at finite
lightlike distance, and it is there that the gauge theory is to be
found.  Note that it has long been understood from the work of
Polyakov that non-critical strings dynamically grow an extra
dimension, so the presence of a five-(plus-five)-dimensional string
theory in the context of a four-dimensional gauge theory is perhaps
not so shocking.  What is astonishing is that the string theory
involved is the well-understood critical superstring, and that it is
believed to be {\it equivalent} to the gauge theory on the boundary.

\begin{figure}
\centering
\epsfxsize=1.6in
\hspace*{0in}\vspace*{0in}
\epsffile{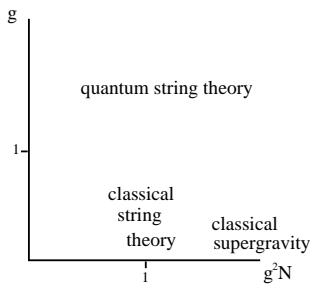}
\caption{The conjectured duality between superstring theory and gauge
theory connects gauge theory to classical superstring theory at small
$g$ and further to classical supergravity at large $g^2N$.}
\label{fig:AdScplg}
\end{figure}
This conjecture is not yet known to be true, but it has been tested in
the large $N$, large $g^2N$ regime, in which the full quantum string
theory reduces to classical (small $g_s$) supergravity (small
curvature), as shown in \FFig{AdScplg}.  In this limit, the symmetries
and operators of the two theories match \cite{maldacon}; and the
baryon operator (which is non-perturbative in string theory since
$N\sim 1/g_s$) has been identified with a D-brane (a soliton)
\cite{ewDbary}.  Furthermore, the Wilson loop of the gauge theory has
been identified with the boundary of a string worldsheet in the larger
space \cite{maldaloop,reyloop}.  This makes it possible to explain how
this conformal \nfour\ SYM theory can be stringy: although the Wilson
loop is the boundary of a string, the string does not live on the
boundary but hangs into the bulk, and so the value of the Wilson loop
as a function of its size {\it depends on the geometry outside the
boundary.}  The difference between area law and perimeter law in
various gauge theories thus is translated into differing spatial
geometries in the string theory duals of those gauge theories.

These ideas have been further extended in a number of directions.  In
particular, it is easy to study finite temperature, and to use the
fact that high-temperature five-dimensional SYM has infrared behavior
equivalent to four-dimensional non-\susic\ YM at strong coupling
\cite{ewThermo,reyThermo,BISYThermo}.  It is straightforward to show
that this strongly coupled theory confines.  It is also possible to
compute its spectrum of glueballs \cite{AdSglubl}.  Remarkably, the
ratios of certain glueball masses match rather well to lattice results
in ordinary QCD away from strong coupling, both in three and four
dimensions.  However, although this is surprising and possibly an
interesting statement about this particular strong-coupling limit,
there is no clear reason for great excitement.  There are many extra
states in the spectrum which do not arise in YM theory; the glueball
mass is not naturally related to the string tension; and there is no
systematic approach toward recovering real YM theory starting from
this limit.

Other extensions include adding matter (which leads to both open and
closed strings), reducing supersymmetry, changing the number of
dimensions, and studying non-conformal theories at zero and finite
temperature.  Various expected phase transitions, such as
deconfinement at high temperature, have been observed.

However, serious obstacles lie in the path of any attempt to apply
these string theory techniques to YM or real QCD, where for any fixed
$N$ the value of $g$ runs such that $g^2N$ is not always large and $g$
is not always small.  Where $g^2N$ is not large, supergravity is
insufficient and string theory is required for the large $N$
limit.  Unfortunately, almost nothing is known about string theory on
an AdS background with Ramond-Ramond fields, even at the classical
level, since the usual world-sheet formulation of string theory cannot
be easily generalized to this case.  Even were this problem solved,
there is no guarantee that the solution will be easy to use.  I cannot
tell you whether these obstacles will be overcome tomorrow, next year,
or in the fourth millenium; but in any case the difficulties are such
that it seems unlikely these approaches will become a quantitative
competitor to lattice gauge theory in the near future.

Nonetheless, it is remarkable that a sensible and definite proposal
for the large $N$ expansion of gauge theory has been made and has
passed some non-trivial tests --- and that it appears to involve
superstring theory!  And we must certainly ask whether in fact
superstring theory can be {\it defined} using gauge theory.

\subsection{A Final Linkage Diagram}

\begin{figure}
\centering
\epsfxsize=2.8in
\hspace*{0in}\vspace*{0in}
\epsffile{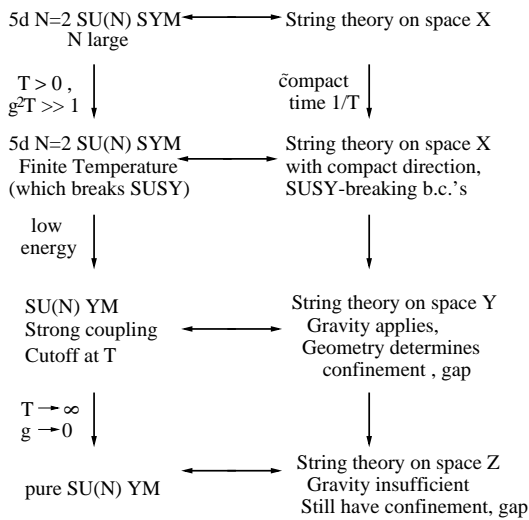}
\caption{The connection of the Maldacena conjecture generalized to
five dimensional \ntwo\ super-Yang-Mills theory to confinement in
Yang-Mills theory.}
\label{fig:linkageC}
\end{figure}

To again bring home how apparently esoteric results on theories with
extended supersymmetry can have implications for real-world physics, I
want to restate using \FFig{linkageC} the connection between QCD and
string theory as presently conjectured and partially understood.  I
must warn the reader that I will be speak rather loosely when
describing this diagram, as this discussion is intended for novices
who want only a rough idea of the physics.  I ask experts to forgive
the obvious misstatements.

Let us consider Euclidean five-dimensional \ntwo\ SYM theory (under
compactification to four dimensions this would become \nfour\ SYM.)
According to the Maldacena conjecture this theory is dual to a
superstring theory which in the large $N$ limit approximately becomes
supergravity on a certain space.  Compactifying the time direction
using periodic boundary conditions for bosons, antiperiodic boundary
conditions for fermions, gives us on the one hand a finite-temperature
version of the SYM theory, and on the other the same supergravity
theory on a space with a compact time direction and
supersymmetry-breaking boundary conditions.  At energies below the
temperature only the bosonic modes invariant under the compact time
remain; the gauge theory appears as four-dimensional and
non-supersymmetric YM, while the gravity is again just the same
theory with the Kaluza-Klein modes removed.  The temperature $T$
serves as a cutoff on the YM theory (since there are many Kaluza-Klein
modes with masses of order $T$) and the coupling $g(T)$ of the YM
theory must be very large at the energy scale $T$ if supergravity is
to be a good approximation.  If we wish to take $T$ to infinity and
the YM coupling $g(T)$ to zero, holding $\Lambda_{YM}$ finite, then
the supergravity regime breaks down and we must include the full
details of the string theory --- a superstring theory with explicit
supersymmetry breaking effects from the finite temperature
construction.

The first  step of this linkage is understood.  The last,
however, is a chasm badly in need of a bridge.

\section{Summary}

I have shown you today that many of the developments in the seemingly
abstract field of supersymmetric field theory have direct or indirect
implications for non-\susic\ YM and QCD.  \none\ SYM is a theory with
properties likely to be shared with non-\susic\ theories.  Predictions
for this theory can be tested on the lattice, and the transition
between SYM and YM is of considerable theoretical interest.  \none\
SQCD, as a function of its gauge group, matter content, and
interactions, shows a wide variety of phenomena, many of which might
appear in non-\susic\ QCD.  Only on the lattice can we search for
these phenomena, some of which might actually play a role in nature.
And the connection between the large $N$ limit of YM and QCD and string
theory, known for many years, now appears to be the tip of a large
iceberg relating gauge theories and gravity/string theories.  Although
quite preliminary, this connection has the greatest potential --- still
far from realized --- to impact our understanding of physical QCD.

   \bibliography{lat98}        
\bibliographystyle{h-physrev}
\end{document}